# Using a Complex Optical Orbital-Angular-Momentum Spectrum to Measure Object Parameters: A Spatial Domain Approach

Guodong Xie[1*,] Haoqian Song[1], Zhe Zhao[1], Giovanni Milione[2], Yongxiong Ren[1],

Cong Liu[1], Runzhou Zhang[1], Changjing Bao[1], Long Li[1],

Zhe Wang[1], Kai Pang[1], Dmitry Starodubov[1], Moshe Tur[3], Alan E. Willner[1]

[1]Department of Electrical Engineering, University of Southern California, Los Angeles, CA 90089, USA

[2]Optical Networking and Sensing Department, NEC Lab. America, Inc., Princeton, NJ 08540, USA

[3]School of Electrical Engineering, Tel Aviv University, Ramat Aviv 69978, Israel

*Corresponding author: guodongx@usc.edu, willner@usc.edu

**Abstract:** Light beams can be characterized by their complex spatial profiles in both intensity and phase. Analogous to time signals, which can be decomposed into multiple orthogonal frequency functions, a light beam can also be decomposed into a set of spatial modes that are taken from an orthogonal basis. Such a decomposition can provide a tool for spatial spectrum analysis, which may allow the stable, accurate and robust extraction of physical object information that may not be readily achievable using traditional approaches. As an example, we measure the opening angle of an object using the complex spectrum of orbital angular momentum (OAM) modes as the basis, achieving a >15 dB signal-to-noise ratio. We find that the dip (i.e., notch) positions of the OAM *intensity* spectrum are dependent on an object's opening angle but independent of the object opening's angular orientation, whereas the slope of the OAM *phase* spectrum is dependent on the object opening's orientation but independent on the opening angle.

# Introduction

When a light beam passes through or is reflected off a physical object or medium, its intensity and phase can be uniquely affected [1]. The information about this object or media could be resolved by investigating the beam's intensity as captured by a camera or its phase as obtained by an interferometer [2]. Analogous to time signals, which can be decomposed into multiple orthogonal frequency functions, a light beam can also be decomposed into a set of spatial modes that are taken from an orthogonal basis [3]. Such a decomposition can provide a tool for spatial spectrum analysis, which may allow the stable, accurate and robust extraction of physical object information that may not be readily achievable using traditional approaches.

An arbitrary beam (*e.g.*, an objet-truncated Gaussian beam) can also be described by its complex spatial spectrum. Such a spectrum is formed by the beam's complex decomposition coefficients over a mutually orthogonal modal basis set [3,4], such as Laguerre-Gauss ($LG_{p,\ell}$, with $p = 0, 1, 2, ...$, as the radial order, and $\ell = 0, \pm1, \pm2, ...$ as the azimuthal order) modes [4-6]. Generally, the spectrum of a beam comprising a single pure mode peaks only at the value corresponding to the order of the mode, while the spectrum of an arbitrary beam could have complex non-zero values for many mode orders.

A subset of LG modes could be the orbital angular momentum (OAM) modes with zero radial index, *e.g.*, $LG_{0,\ell}$ [5]. An OAM mode has a phase front of $\exp(j\ell\phi)$, which 'twists' in a helical fashion as it propagates [5], where $\ell$ is also referred to as the OAM order, and $\phi$ is the azimuthal angle. OAM modes are incomplete in the radial coordinate but they could form a complete orthogonal basis in the azimuthal coordinate. Therefore, when they are used as the orthogonal basis on which to characterize an object-truncated beam, the coefficient of various orders of OAM modes may represent the various azimuthal properties of the object [3,7-10].

The properties of OAM modes have included the following: (1) The phase-change rate of an OAM mode is proportional to its order, meaning that a larger-order OAM beam has a smaller phase-change spatial periodicity [9,10]. (2) The complex OAM spectrum of an arbitrary beam could form a Fourier pair with its spatial-intensity distribution in the azimuthal direction [3,9,10]; (3) The intensity of an OAM mode is circularly symmetric, which means the intensity of a beam's OAM spectrum is generally rotation insensitive [5]; and (4) An OAM mode is relatively stable in a homogeneous medium, which indicates that the amount of OAM of a beam could be constant during free-space propagation regardless of the beam diffraction [1,2].

The use of structured beams has recently been investigated in imaging [4,11-25], sensing [4,11,26-36], and communications [4,11,37,38] and other applications [4,11,39], including using OAM basis for single-frequency imaging and sensing in the radio frequency domain [24] and the quantum domain [20]. It may be desirable to design a complex OAM spectrum analyzer in the classical optical domain and explore its potential to provide information that could not be readily obtained using the traditional approach.

In this paper, we demonstrate the use of OAM-based complex spectral analysis in the classical optical domain to measure object parameters. The OAM intensity spectrum (amplitude of the complex decomposition coefficients) and OAM phase spectrum (angle of the complex decomposition coefficients) of an object-truncated Gaussian beam are measured by carefully designing a phase mask in the OAM spectral analyzer. Using the measured complex OAM signature, we could identify the object's relative shape information from the OAM intensity spectrum and its position information from the OAM phase spectrum, where it is difficult for the traditional approach to measure. Specifically, we explore the potential of a complex OAM spectrum based system to detect the opening angle and orientation of a sector-shaped object, achieving a > 15 dB signal to noise ratio. Our results show the followings: (1) The OAM intensity spectrum is dependent on the opening angle of the object but insensitive to its orientation; (2) The OAM intensity spectrum is relatively insensitive to free-space beam propagation in the lab, where the measured OAM intensity spectrum is nearly identical when the distance from the object to the OAM spectrum analyzer varies between 0 and 104 cm; and (3) The OAM phase spectrum is dependent on the orientation of the object but insensitive to its opening angle.

Figure 1 shows the concept of using an OAM spectral analyzer to measure object parameters. To measure an object's parameters, including shape, thickness, and temperature, a light beam (Gaussian beam, OAM beam, or other beams) could be shone onto an object, and by investigating this beam after the object-truncation, the parameter of interest could be retrieved. In this work, we use the measurement of the opening angle and the opening's orientation of an object as an example. When a camera is used to capture the intensity of the beam truncated by the object, the following problems might arise: (1) Due to the diffraction caused by object truncation and beam propagation, the image may become too blurry to quantify the opening angle of the object, as shown in the intensity display in Figure 1c, especially when the object is not focused on the camera; (2) If the object is rotating at a relatively high speed, the camera may not be able to capture a steady image; (3) A camera is usually a multi-pixel device, which may cost more than a single-pixel device; and (4) A camera usually could not capture the phase information of the object-truncated light.

However, the complex OAM spectrum based approach could potentially determine the opening angle and opening's orientation of an object without suffering the above-mentioned issues. This approach is based on the following facts: (1) Object truncation of the probe beam could change its OAM spectrum, and the OAM spectrum before and after the object truncation differ sufficiently that the 'truncation' could be identified; (2) The complex OAM spectrum of a beam forms a Fourier pair with its spatial distribution in the azimuthal direction[7-10], in which an opening on the object would lead to a Sinc function in its OAM intensity spectrum; (3) The light, propagating onward from the truncated object, maintains the powers of its constituent OAM modes so that its OAM intensity spectrum is unaffected by the radially symmetric diffraction kernel[2]; and (4) The phase-change rate of an OAM is proportional to its order; therefore, rotating an object may cause different phase delays to different components on the OAM phase spectrum.

**Results**

Figure 2a shows the experimental setup. A spatial light modulator (SLM-1) is used to generate the desired probe beam with a certain OAM order. We generally use a Gaussian beam ($\ell = 0$) as the probe, and we also show the cases when various orders of OAM beams are used as the probes to measure object parameters. SLM-2 is used to emulate objects with various parameters (opening angles, orientations and numbers of opening slot). Figure 2b shows the properties of these objects. We made the following assumptions: (1) The object has a larger size than the probe beam; (2) The object has a sector-shaped opening that is characterized by θ; and (3) The orientation of the opening is defined as the angle between the left edge of the slot opening and the y-axis, which is characterized by $\delta$.

The beam truncated by the 'object' is then collected by a complex OAM spectrum analyzer, which is composed of SLM-3, a subsystem coupling light from free space to single mode fiber (SMF), and a power monitor. To obtain a complex OAM spectrum, we measure the OAM intensity spectrum and the OAM phase spectrum separately. To measure the OAM intensity spectrum, SLM-3 is loaded with various spiral phase patterns, so that the decomposition of incoming light onto various OAM orders is measured sequentially and together forms an OAM intensity spectrum (see Methods section for details). To measure the OAM phase spectrum, we measure the relative phase between OAM $\ell$ ($\ell = -8, -7, ..., +7, +8$) and OAM 0 (*i.e.*, Gaussian mode) sequentially. For each phase measurement, four different specially designed patterns are loaded on SLM-3 sequentially, and together the four intensities measured by the power monitor enable calculation of the phase (See Methods section for details). In this experiment, the distance between the object (SLM-2) and the OAM spectrum analyzer (SLM-3) varies from 0 to 35, 52 and 104 cm. We note that for 0 distance, SLM-2 serves only as a mirror, and the 'object' emulation pattern and OAM spectrum analyzer pattern are combined and loaded on SLM-3.

We first use the the complex OAM spectrum based approach to measure the opening angle of an object. As Fig. 3a shows, the sample object has an opening angle of $2\pi/3$ and is placed in the propagation path of a Gaussian probe beam. For comparison, we also measure the image of the object-truncated light with a lens-less camera. As Fig. 3b shows, it is difficult to determine the opening angle of the object due to blurring. However, when we measure the OAM intensity spectrum, we observe several dips in the OAM spectrum, as shown in the Fig. 3c. The power dips appear on OAM orders $|\ell| = 3N$ ($N$ is a non-zero integer), indicating that the object has an opening angle of $2\pi/3$. When an object, azimuthally opened with an angle of $\theta$, truncates an OAM probe with order $\ell_1$ (here $\ell_1 = 0$), the probe's spectral OAM component of order $\ell_2$ vanishes when $\ell_1 - \ell_2 = 2\pi N/\theta$ [7,8,9,13]. This is due to the 'vanishing effect' of the overlap integral between $\ell_1$ and $\ell_2$ over an azimuthal angle of $\theta$, that is $\int_0^\infty \int_0^\theta u(r,\phi,\ell_1) u^*(r,\phi,\ell_2)\, d\phi dr = 0$, when $(\ell_1 - \ell_2)\theta = 2N\pi$, where $u(.)$ is the electrical field. The power difference between a dip and its neighbors has a difference of > 15 dB, indicating that our approach has a high signal-to-noise ratio (SNR).

When we rotate the object in Fig. 3a with various angles (0, $\pi/4$, $\pi/2$, $3\pi/4$, $\pi$, $5\pi/4$, $3\pi/2$ and $7\pi/4$) along its center (corresponding to the states 1 to 8 in Fig. 3b), the measured OAM spectrum is almost unchanged, as Fig. 3c shows. We believe this is due to the circular symmetry of the intensity of an OAM beam. Therefore, our approach could potentially be used to determine the opening angle of the object even when the object is rotating.

Furthermore, we verify the relationship between object's opening angle and the OAM intensity spectrum of the beam truncated by the object. As Fig. 3d shows, when the object's opening angle is modified to $2\pi/4$, dips are observed on the modes with $|\ell| = 4N$ (e.g., $\ell = -8, -4, +4, +8$). When the opening angle is $2\pi/5$, dips are observed on the modes with $|\ell| = 5N$ (e.g., $\ell = -5, +5$). To explore the relationship between the object's opening angle and the first dip position that appears, we simulate a more general case, as shown in Fig. 3e. The first dip position is inversely proportional to the object's opening angle.

The object might have more than one opening slot, as shown in Fig. 3f. Here, we explore three different objects with one, two and three slots, each of which has an opening angle of $\theta = 2\pi/6$. As Fig. 3g shows, the images measured using a camera are still blurred. However, the OAM spectrum shows clear characteristics that help determine both the object's opening angle and its slot number. As Fig. 3h shows, when there are two slots, peaks appear at every other mode, while when there are three slots, peaks appear every three modes. This peak periodicity indicates the number of opening slots on the object. No matter how many slots the object has, deep dips show up on OAM +6 and OAM-6, indicating that the size of the opening is $2\pi/6$.

As Fig. 3i shows, the OAM spectrum varies within a small range when the spectrum analyzer is placed at various distances (0 cm, 35 cm, 52 cm and 104 cm) from the object. This is because the kernel of the diffraction (the effect of propagating free space) is proportional to the phase-changing direction of the OAM beam. However, we observe some power fluctuation at various distances. We believe this is because our system ignores LG modes with non-zero radial orders ($p \neq 0$). When the beam is truncated by an object and propagates in free space, higher-radial-order LG modes may arise whose power could not be collected by our OAM spectrum analyzer since it contains a free space to SMF coupling subsystem.

We discuss the case when a Gaussian beam is used as the pilot to measure the object parameter. An OAM beam could also be used as the pilot beam. As Fig. 3j shows, when various orders of OAM beams are used sequentially as probes, the magnitudes of the measured OAM spectral components form a circulant matrix. Such a matrix could provide multiple copies of information for measuring object parameters, thus potentially ensuring higher accuracy by averaging the various measurements.

Besides the OAM intensity spectrum, the resultant OAM modes may also have various phases. Here, we sequentially measured the relative phase between OAM $\ell$ and OAM 0 for $\ell = -8, -7, ..., +7, +8$, to get an OAM phase spectrum. In general, to measure the phase difference between OAM $\ell_1$ and OAM $\ell_2$, we load phase masks $T_0, T_{45}, T_{90}$ and $T_{135}$ on SLM-3, where:

$$\begin{cases} \exp(jT_0) = \exp(j\ell_1\phi) + \exp(j\ell_2\phi) \\ \exp(jT_{45}) = \exp(j\ell_1\phi) + j\exp(j\ell_2\phi) \\ \exp(jT_{90}) = \exp(j\ell_1\phi) - \exp(j\ell_2\phi) \\ \exp(jT_{135}) = \exp(j\ell_1\phi) - j\exp(j\ell_2\phi) \end{cases} \quad (1)$$

We define the intensities measured by the power monitor when loading each phase mask, $T_0, T_{45}, T_{90}$ and $T_{135}$, on SLM-3 as $I_0, I_{45}, I_{90}$ and $I_{135}$, respectively. Then, the relative phase difference between the two OAM modes is given by (See Methods section for details)[29,36,40]:

$$\Delta\varphi = \text{atan}((I_0 - I_{90})/(I_{45} - I_{135})) \quad (2)$$

We note that reports have shown the complex OAM spectrum measurement using an interferometer[34,35], which might be sensitive to the setup vibration since its two branches usually have different optical paths. However, the approach in this work is based on non-interference[29,36,40] and may provide a more stable phase measurement.

We measure the OAM phase spectrum of the object-truncated beam when the object has an opening angle of $2\pi/3$ but various orientation angles $\delta$. As Figures 4a and 4b show, the orientation angle is closely related to the slope of the OAM phase spectrum. When the orientation angle of the object is $\delta$, the slope of the measured phase is approximately $\delta$, where a different sign of $\delta$ indicates that the object is oriented in a different direction relative to the y-axis. We believe this is because different OAM modes have different phase-change rates. Rotating the object by a certain angle may cause different phase delays for different OAM modes, and such phase delays are proportional to the OAM orders (see Fig. 4c).

We also fix the orientation of the object to be $\pi/8$. When we change the object's opening angle from $2\pi/3$ to $2\pi/4$, $2\pi/5$ and $2\pi/6$, the slope of the OAM spectrum is nearly constant, as shown in Fig. 4d, indicating that the OAM phase spectrum is insensitive to the object's opening angle.

Our phase measurement is based on calculating an Arctan function (Equation 2), which means we could only resolve a phase between $-\pi/2$ and $\pi/2$. This might cause phase-measurement degradation when the object has a large orientation angle ($\delta$). As Figure 4e shows, when $\delta = \pi/4$, the measured phase shows periodicity every three or four OAM modes, making it difficult to determine the slope of the phase. However, this could be compensated for by calculating a pre-estimate $\delta_0$. Using a pre-estimate, the phase masks loaded to SLM-3 are calculated by:

$$\begin{cases} \exp(jT_0) = \exp(j\ell_1\phi) + \exp(j\ell_2\phi + j\ell_2\delta_0) \\ \exp(jT_{45}) = \exp(j\ell_1\phi) + j\exp(j\ell_2\phi + j\ell_2\delta_0) \\ \exp(jT_{90}) = \exp(j\ell_1\phi) - \exp(j\ell_2\phi + j\ell_2\delta_0) \\ \exp(jT_{135}) = \exp(j\ell_1\phi) - \exp(j\ell_2\phi + j\ell_2\delta_0) \end{cases} \quad (3)$$

Figure 4f shows the OAM phase spectrum measurement for an object with an opening angle of $2\pi/3$ and an orientation angle of $\delta = \pi/4$, when we pre-estimate $\delta_0 = \pi/8$. Using this pre-estimate, the slope of the measured phase exhibits less fluctuation.

**Discussion**

There are several issues that are valuable to consider when using a beam's complex spectrum for object parameter measurement.

The approach outlined in this paper can be extended to provide higher resolution than we have experimentally shown. As one example, the object could have an opening angle of 2mπ/N instead of 2π/N, where m is an integer. Figure 5a shows the cases when N = 5, and m = 1, 2, 3 and 4. We find that the position of the dip is determined by N (see Characteristic B), while the power of the transmitted probe beam (the peak or the ratio of peak over its neighbor) reflects the relative information of m (see Characteristic A) because larger m allows more received power.

As a second example, the resolution of the measured opening angle could be further improved by increasing the number of OAM modes that have been investigated. As a proof-of-concept, we investigate the power and phase of OAM beams with orders of from -8 to +8. Evidently more modes could be used to improve the resolution of the measurement as indicated by the simulation results in Figure 3e.

OAM modes are used as basis for measuring the object opening angle and object opening's orientation. Different basis such as LG modes or Hermite-Gauss (HG) mode might also be used, depending on the specific parameters to be measured and the unique properties of the modes.

**Method**

To measure the OAM intensity spectrum of an object-truncated beam, we load the various spiral phases ($-\ell\phi$) on SLM-3 sequentially. Each pattern down-converts a specific OAM component (OAM+$\ell$) to a Gaussian-like beam that could be collected by the single mode fiber (SMF), while the other component is converted to non-zero OAM modes, which could not be collected by the SMF. When $\ell$ varies between -8 and 8, an OAM spectrum is obtained.

Figure 6 shows the principle for measuring the phase between OAM $\ell_1$ and OAM $\ell_2$. In general, measuring the phase difference between two OAM modes is analogous to measuring the relative phase difference between the left-circularly polarized component and the right-circularly polarized component of a light beam in Stokes polarimetry[26]. First, we define a light beam using left-circularly polarized and right-circularly polarized light as the basis for $E = E_l \hat{l} + E_r \hat{r}$, where $E$ is the electrical field of the beam and $E_l$ and $E_r$ denote the components of the beam that are polarized left- and right-circularly, respectively. We could have x- and y-polarized light as another basis for representing the light beam $E = E_x\hat{x} + E_y\hat{y}$,

$$\begin{cases} E_x = E_l + E_r \\ E_y = E_l - E_r \end{cases} \tag{4}$$

where $E_x$ and $E_y$ denote the x- and y-polarized component of the beam, respectively. We could also have another basis, which is light that is polarized 45-degrees ($E_w$) and 135-degrees ($E_v$), which obeys $E = E_w\hat{w} + E_v\hat{v}$, where:

$$\begin{cases} E_w = E_l + jE_r \\ E_v = E_l - jE_r \end{cases} \tag{5}$$

We use $I_0, I_{45}, I_{90}$ and $I_{135}$ to characterize the intensity of $E_x, E_y, E_w$ and $E_v$, respectively. The $I_0, I_{45}, I_{90}$ and $I_{135}$ of a light could be measured when we have a linear polarizer rotated 0, 45, 90 and 135 degrees, respectively, with respect to its transmission axis. Then, the relative phase difference between components of the light that are polarized left- and right-circularly is given by [29,36,40]:

$$\Delta\varphi = \operatorname{atan}\left((I_0 - I_{90})/(I_{45} - I_{135})\right) \tag{6}$$

Analogously, when we replace the original basis (beams that are polarized left- and right-circularly) with two OAM modes (with the orders of $\ell_1$ and $\ell_2$), we are able to measure the relative phase difference between the two OAM modes. We load phase masks $T_0, T_{45}, T_{90}$ and $T_{135}$, which are analogous to a linear polarizer rotated 0, 45, 90 and 135 degrees, respectively, on SLM-3, where $T_0, T_{45}, T_{90}$ and $T_{135}$ obey:

$$\begin{cases} \exp(jT_0) = \exp(j\ell_1\phi) + \exp(j\ell_2\phi) \\ \exp(jT_{45}) = \exp(j\ell_1\phi) + j\exp(j\ell_2\phi) \\ \exp(jT_{90}) = \exp(j\ell_1\phi) - \exp(j\ell_2\phi) \\ \exp(jT_{135}) = \exp(j\ell_1\phi) - j\exp(j\ell_2\phi) \end{cases} \tag{7}$$

Similarly, we define the intensities measured by the power monitor when loading each phase mask, $T_0, T_{45}, T_{90}$ and $T_{135}$, on SLM-3 as $I_0, I_{45}, I_{90}$ and $I_{135}$, respectively. Then, the relative phase difference between the two OAM modes is given by the same Equation (6). Figure 6 shows the phase profiles for OAM 0 and OAM+3 and the $T_0, T_{45}, T_{90}$ and $T_{135}$ phase patterns used to measure their relative phases.

**Acknowledgement**

We acknowledge the supports from National Science Foundation (NSF) (ECCS-1509965) and the Vannevar Bush Faculty Fellowship program sponsored by the Basic Research Office of the Assistant Secretary of Defense for Research and Engineering and funded by the Office of Naval Research (ONR) (N00014-16-1-2813).

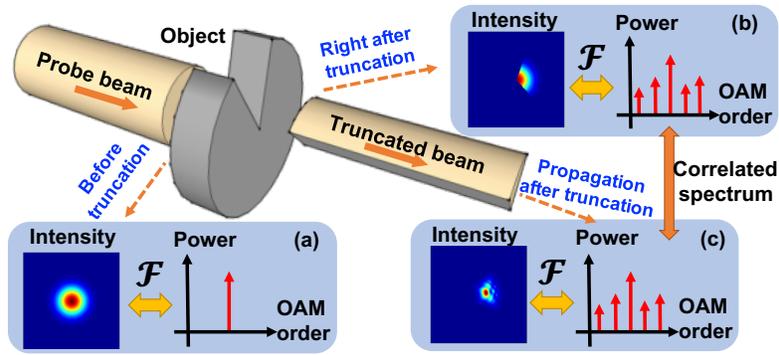

**Figure 1.** The concept of using the OAM spectrum to measure an object's parameters. The beam's intensity profile and OAM spectrum (a) before object truncation, (b) right after object truncation, and (c) some distance after object truncation.

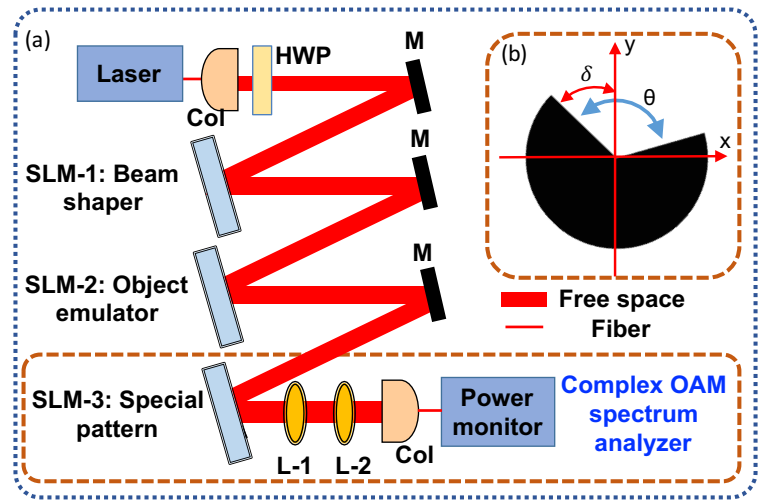

**Figure 2.** (a) Experimental setup. The SLM-3, L-1, L-2, collimator and power monitor form the complex OAM spectrum analyser. Col: collimator, HWP: half-wave plate, M: mirror, SLM: spatial light modulator, L: lens. (b) The shape and position of the object. θ: opening angle of the object; δ: orientation of the object relative to the y-axis.

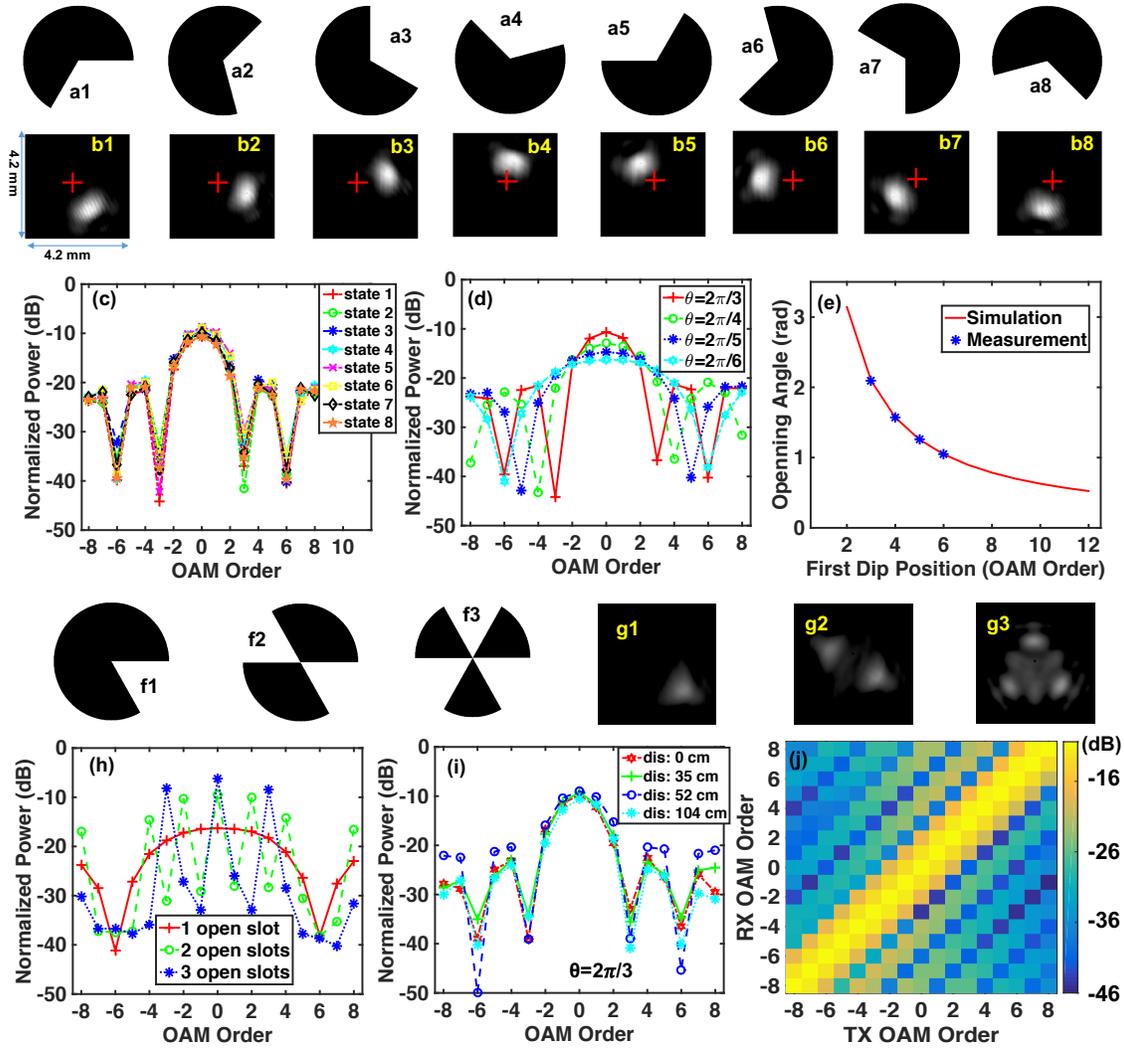

**Figure 3.** (a1–a8) Various orientations (states) for an object having an opening angle of $2\pi/3$. (b1–b8) The image of the light beam truncated by the objects in (a1–a8), respectively. (c) The OAM intensity spectrum measured for the light truncated by the objects in (a1–a8). (d) The OAM intensity spectrum measured for the light truncated by objects having various opening angles. (e) The relationship between the opening angles and the first-dip position in the OAM intensity spectrum. (f1–f3) Objects having 1, 2 and 3 slots, each of which has an opening angle of $2\pi/6$. (g1–g3) The images of the light beam truncated by the objects in (f1–f3). (h) The OAM intensity spectrum measured for the light truncated by the objects in (f1–f3). (i) The OAM intensity spectrum measured for the light truncated by an object having an opening angle of $2\pi/3$ when the distance from the object to the OAM spectrum

analyser is 0 cm, 35 cm, 52 cm and 104 cm. (j) The OAM intensity spectrum matrix measured when various OAM beams are used as the probe. The distance from the object to the OAM spectrum analyser is 0 in c, d, e and h, j.

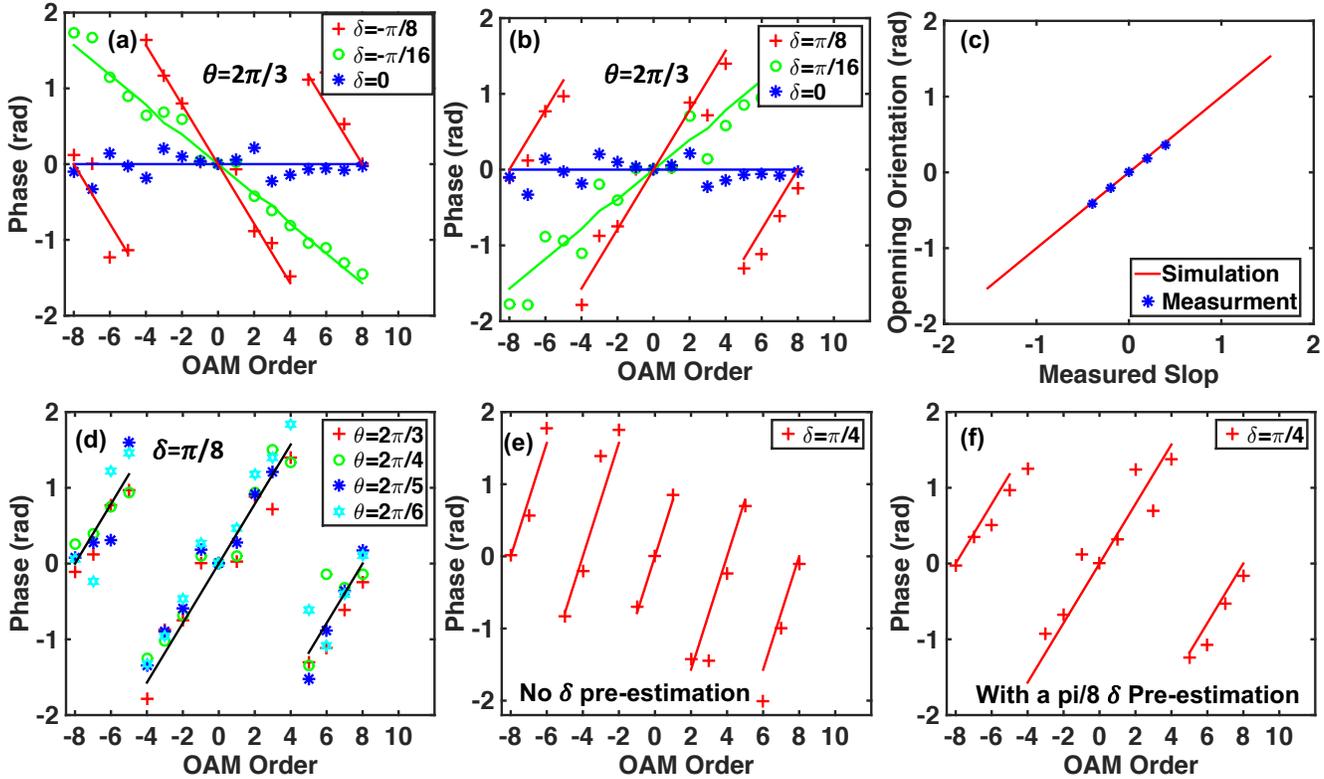

**Figure 4.** (a) The OAM phase spectrum measured for the light truncated by an object having an opening angle of $2\pi/3$ but various negative orientation angles. (b) The OAM phase spectrum measured for the light truncated by an object having an opening angle of $2\pi/3$ but various positive orientation angles. (c) The relationship between the orientation angle of the object and the slope of the OAM phase specturm. (d) The OAM phase spectrum measured for the light truncated by an object having various opening angles and an orientation angle of $\pi/8$. (e, f) The OAM phase spectrum measured for the light truncated by an object having an opening angle of $2\pi/3$ and an orientation angle of $\pi/4$. (e) No pre-estimate for $\delta$. (f) Pre-estimate of $\delta = \pi/8$. In this Figure, the distance from the object to the OAM spectrum analyser is 0. The lines show the simulation results, and the symbols show the experimental measurements. In the measurement, the data is calculated by a Arctan function which is between $-\pi/2$ and $\pi/2$, and we may also add a $\pi$ or $-\pi$ phase shift to the measurement for the convenience of slop calculation.

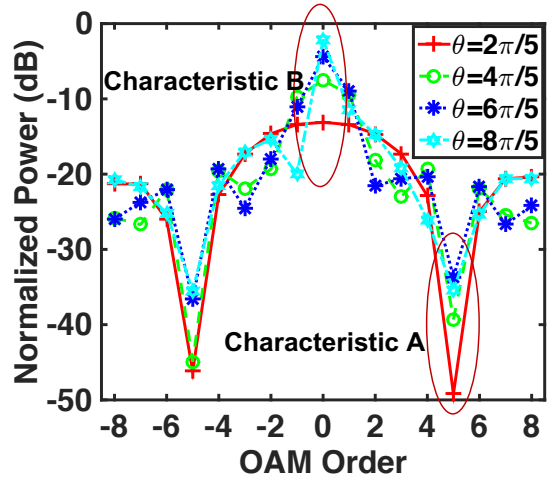

**Figure 5.** The OAM intensity spectrum measured for the light truncated by an object having an opening angle of 2mπ/5, where m = 1, 2, 3 and 4.

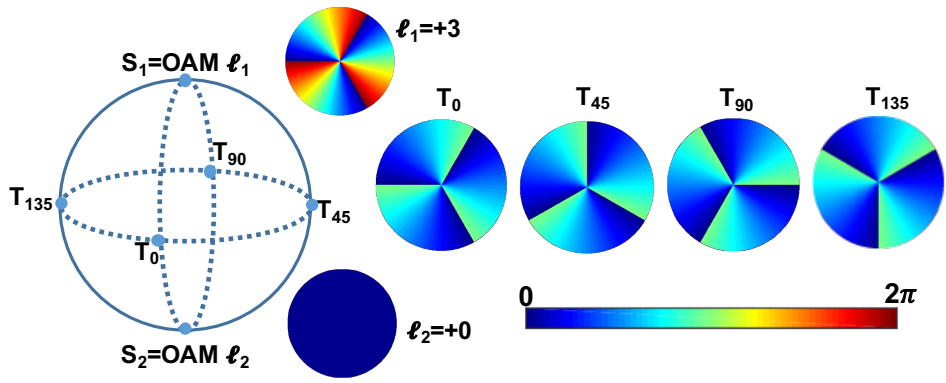

**Figure 6.** Principle used to measure the OAM phase spectrum. $T_0$, $T_{45}$, $T_{90}$ and $T_{135}$ are calculated according to $\ell_1$ and $\ell_2$ and are loaded to SLM-3 sequentially to measure the phases between OAM $\ell_1$ and $\ell_2$.